\documentclass[sigconf,nonacm]{acmart}
\usepackage{booktabs}
\usepackage{array}
\usepackage{graphicx}
\usepackage{tikz}
\usetikzlibrary{positioning,arrows.meta,backgrounds,fit,calc}
\usepackage{pgfplots}
\pgfplotsset{compat=1.18}
\graphicspath{{figures/}}

\definecolor{figink}{HTML}{1B2733}
\definecolor{figmute}{HTML}{65788A}
\definecolor{figline}{HTML}{C6D2DC}
\definecolor{figlink}{HTML}{377DD8}
\definecolor{figpath}{HTML}{7A5CC0}
\definecolor{figreal}{HTML}{D98A00}
\definecolor{figgood}{HTML}{1F9D57}
\definecolor{figbad}{HTML}{D23B3B}

\begin{document}

\title[Paired measurements and fault attribution from a month-long WiFi
probe deployment]{A Remote WiFi Performance Tool: Paired Measurements and
Fault Attribution from a Month-Long Unattended Deployment}

\author{Selim Gul}
\affiliation{%
  \institution{University of Liverpool}
  \city{Liverpool}
  \country{United Kingdom}}
\email{gul.selim@outlook.com}

\begin{abstract}
A throughput figure does not show how well an application will perform
on a WiFi network: whether a video will stall, or where slowness comes
from. This matters most on public and shared networks, whose users
control neither access point nor uplink. We present a tool that measures
applications from inside the network, and report where its fault
diagnosis failed.

In the developed architecture, a Raspberry~Pi probe joins the network as
a client. It regularly loads a web page, starts a video, fetches email
and downloads a file, and every 10~s measures round-trip time and packet
loss to five destinations. Web, video and download each run against a
local reference server, an identical cloud copy and the real service,
and the pattern of the three results is read, under stated assumptions,
as evidence for a fault in the wireless link, the uplink or the service.

The probe ran unattended for 32 days in student accommodation, where
traffic between clients is blocked, so no local reference could run and
round trips to the default gateway stood in for the link. Over the last
10.4 days, the round trip to a public address was a median of 7.75~ms
above the gateway's in the same 10~s; when the gateway's exceeded its
own 95th percentile, the other's exceeded its own in 92\% of cases,
against 0.4\% otherwise. A 39-minute loss of off-site reachability read
as missing data to the deployed code and, after corrections, as a fault
beyond the gateway.

Five injected impairments and a control were scored against results
fixed beforehand. They exposed three more defects; once fixed, the
control and three impairments were diagnosed as expected, and the two
misses are limits of fixed thresholds.

Monitoring from a WiFi client proved practical; diagnosis by fixed
thresholds was not reliable. Eight times the tool itself created or hid
an apparent fault, through errors of collection, aggregation or
presentation.

\end{abstract}

\maketitle

\section{Introduction}
\label{sec:intro}

The usual test of a WiFi network is a speed test, which reports upload
and download throughput; some now add latency, jitter and packet loss
and turn one run into scores for streaming, gaming and
calls~\cite{cloudflare-aim}. A throughput figure does not say whether a
video will start without stalling or whether a page will load promptly.
One run from one device also does not say where a problem lies: in the
wireless link, in the building's connection to the internet (its
\emph{uplink}), or in the service being used. Throughput and experience
are known to differ. Across a large video platform, time spent
rebuffering affected how long people kept watching more than the bitrate
they received~\cite{dobrian2011video}, and a page has not loaded when
its bytes arrive but when the browser has parsed and run
them~\cite{wang2013wprof}. A single test also misses change over time. A
network that is fine at a quiet hour can be unusable at a busy one,
which is why studies of home networks use devices that measure a link
continuously for
weeks~\cite{sundaresan2011broadband,sundaresan2014bismark}. The problem
is sharpest on public and other shared WiFi, in a caf\'e, a library or a
hall of residence, where the user controls neither the access point nor
the uplink and cannot tell whether the link, the uplink or the service
is at fault.

This paper describes a tool that measures one WiFi network from the
inside, and reports what a month of unattended operation showed about
the network and, more often, about the tool. The probe is a
Raspberry~Pi that joins the network as an ordinary client. On a schedule
it performs four tasks as a user would, called \emph{workloads}: it
loads a web page in a real browser, fetches a video as a player would,
fetches mail and downloads a file. Every ten seconds it also runs a
\emph{baseline}, measuring round-trip time and packet loss to five
destinations, among them the network's \emph{default gateway} (the
router the probe is attached to) and two well-known public addresses,
the \emph{public anchors}. The central idea is that the web, video and
download workloads each run against three destinations, or
\emph{endpoints}: a reference server wired to the same local network, a
cloud server holding identical content, and the real service. All three
paths cross the wireless link, and the cloud and real paths also share
the building's uplink before they diverge. Each result is classed as
good, degraded or poor, its \emph{verdict}, and the pattern of the three
verdicts is evidence for where a fault lies
(Section~\ref{sec:design}). Naming the part of the path at fault is
called \emph{attribution}. Two dashboards present the results, one of
which answers a resident's question: \emph{is my WiFi supporting what I
do, and if not, where is the slowness?}

The system ran for 32 days in a student accommodation building in
Liverpool in July and August 2026. The site's network does not let two
client devices reach each other, so no local reference could be run
there (Section~\ref{sec:impl}). The comparison therefore had two
endpoints, not three, and the wireless link was judged by round trips to
the default gateway.

None of the ingredients are new on their own: small probes, browser
workloads, comparison with the gateway and rule-based diagnosis are all
established. Most published work measures access networks, the last link
between a user and the internet, from the gateway, from the access point
or from platforms of dedicated probes, and the research tools that run
from a client inside the network run network tests, not application
workloads. Commercial endpoint monitoring does combine application tests
with wireless and gateway data from an agent on the user's machine, in
closed products~\cite{thousandeyes-endpoint}
(Section~\ref{sec:related}). What this paper adds is an open
implementation of one combination, a client inside the WiFi network
running application workloads against a reference copy and the real
service, and a record of how its diagnosis fared against a real fault
and injected ones, failures included. This paper presents a case study
of one site. The contributions include:

\begin{itemize}
\item \textbf{A three-way endpoint comparison from a WiFi client.} The
  web, video and download workloads run against a wired local reference,
  a cloud copy of the same content and the real service, and the three
  verdicts are read as a pattern. No time measured at one endpoint is
  subtracted from another's. The code is public
  (Section~\ref{sec:design}).
\item \textbf{A paired latency comparison, and a correction.} The
  gateway and a public anchor are measured within the same ten seconds,
  so their round trips can be compared measurement by measurement. Over
  10.4 days the anchor's was a median of 7.75~ms above the gateway's,
  and when the gateway's was above its own 95th percentile the anchor's
  was above its own in 92\% of cases, against 0.4\% otherwise. The two
  move together; this does not show how delay divides along the path.
  The dissertation this paper condenses~\cite{gul2026dissertation}
  reported the gateway as 29\% of the median path and as the source of
  most of its variability. Both claims compared different periods and
  probe settings, and neither survives the paired analysis
  (Section~\ref{sec:deployment}).
\item \textbf{An attribution rule tested on a real fault and on injected
  ones.} For 39 minutes on 18 August 2026 everything beyond the gateway
  was unreachable while the gateway kept answering. The stored records
  show that the rule as deployed would have reported missing data; with
  two defects corrected, the same records give a reachable local network
  and a fault beyond it (Section~\ref{sec:outage}). Five impairments and
  a control with none were then applied with the Linux traffic shaper
  \texttt{tc netem} and scored against verdicts written down beforehand.
  They exposed three more defects. After correction the control and
  three of the five impairments are attributed as expected, on the same
  scenarios, and the two misses are limits of fixed thresholds
  (Section~\ref{sec:faultinjection}).
\item \textbf{A catalogue of eight measurement artefacts.} A measurement
  artefact is a result produced by the tool, not by the network. Eight
  times the project reported a network problem that did not exist, or
  missed one that did. The lesson is that a tool must be validated both
  in how it collects data and in how it presents them
  (Section~\ref{sec:artefacts}).
\end{itemize}

Section~\ref{sec:design} gives the design and Section~\ref{sec:impl}
the implementation and what could not be built at the site.
Section~\ref{sec:deployment} reports the deployment,
Sections~\ref{sec:outage} to~\ref{sec:artefacts} the outage, the fault
injection and the artefacts. Section~\ref{sec:related} places the work,
Section~\ref{sec:limitations} states its limits and
Section~\ref{sec:conclusion} concludes. The code is public under the
MIT licence.\footnote{\url{https://github.com/Selimgul14/network-quality-analysis}}

\section{Design}
\label{sec:design}

The system has four parts: the probe, a local reference server, a cloud
backend and two dashboards. The probe runs workloads and produces
records, the backend stores and queries them, and the attribution code
turns them into a diagnosis. The parts share only the measurement record
of Section~\ref{sec:contract}, which let them be built and changed
independently.

\paragraph{Terms.}
Four more words are used in a fixed sense. A workload's \emph{headline
metric} is the one figure it is judged by: page load time, video startup
time, mail fetch time or download throughput. A \emph{segment} is one of
the three parts of the path that the tool can name: the wireless link,
the internet path beyond it (the building's uplink and what follows) and
the third-party service. One ten-second round of the baseline is a
\emph{cycle}, and each kind of destination it probes is a
\emph{destination class}.

\begin{figure*}[t]
\centering
\begin{tikzpicture}[
  font=\small,
  box/.style={rounded corners=3pt, draw=figline, fill=white,
              inner sep=5pt, align=center, text=figink},
  seg/.style={rounded corners=9pt, draw=#1, fill=#1!10, text=figink,
              inner xsep=7pt, inner ysep=4pt, font=\footnotesize},
  dest/.style={box, minimum width=2.7cm, minimum height=0.95cm},
  flow/.style={-{Stealth[length=4pt]}, draw=figmute, thin},
  wire/.style={draw=figmute, thin},
]

\node[box, minimum width=2.0cm, minimum height=1.9cm] (probe)
  {\textbf{Probe}\\[1pt]\footnotesize\color{figmute}one workload,\\
   \footnotesize\color{figmute}three targets};

\node[seg=figlink, right=0.5cm of probe.east] (wifi) {wireless link};
\coordinate (j1) at ($(wifi.east)+(0.45cm,0)$);

\node[dest, dashed, anchor=west] (local) at ($(j1)+(0.5cm,1.55cm)$)
  {Local reference\\[-1pt]\footnotesize\color{figmute}identical content;\\[-1pt]
   \footnotesize\color{figmute}not run at the site};

\node[seg=figpath, anchor=west] (uplink) at ($(j1)+(0.5cm,-0.775cm)$) {building uplink};
\coordinate (j2) at ($(uplink.east)+(0.45cm,0)$);

\node[seg=figpath, anchor=west] (cpath) at ($(j2)+(0.5cm,0.775cm)$) {path to cloud host};
\node[dest, right=0.5cm of cpath] (cloud)
  {Cloud reference\\[-1pt]\footnotesize\color{figmute}identical content};

\node[seg=figpath, anchor=west] (rpath) at ($(j2)+(0.5cm,-0.775cm)$) {path to real host};
\node[seg=figreal, right=3pt of rpath] (svc) {the service};
\node[dest, right=0.5cm of svc] (real)
  {Real service\\[-1pt]\footnotesize\color{figmute}e.g.\ a news site};

\draw[wire] (probe.east) -- (wifi.west);
\draw[wire] (wifi.east) -- (j1);
\draw[flow] (j1) |- (local.west);
\draw[flow] (j1) |- (uplink.west);
\draw[wire] (uplink.east) -- (j2);
\draw[flow] (j2) |- (cpath.west);
\draw[flow] (j2) |- (rpath.west);
\draw[flow] (cpath.east) -- (cloud.west);
\draw[flow] (svc.east) -- (real.west);

\node[below=1.0cm of probe.south, anchor=north west, xshift=-1.0cm,
      align=left, font=\footnotesize, text=figink] (rule) {%
  \textbf{Reading the three verdicts together}\\[3pt]
  \textcolor{figlink}{$\blacksquare$} all three slow \quad$\rightarrow$\quad the wireless link, or the probe itself\\
  \textcolor{figpath}{$\blacksquare$} local fine, other two slow \quad$\rightarrow$\quad the uplink, or the paths beyond it\\
  \textcolor{figreal}{$\blacksquare$} only the real service slow \quad$\rightarrow$\quad that service, or its path\\[4pt]
  \footnotesize\color{figmute}The cloud and real paths share the uplink and then diverge; neither contains the other.\\
  \footnotesize\color{figmute}A pattern is evidence for a class of fault, not a measurement of one segment.};
\end{tikzpicture}
\caption{The three-way endpoint split. All three paths cross the wireless
link, and the cloud and real paths also share the building's uplink and
then diverge; the attribution logic
of Section~\ref{sec:attribution} reads the pattern of the three verdicts
as evidence for a fault class, under the assumptions stated in
Section~\ref{sec:threeway}. The local
reference is drawn dashed because the deployment site enforced client
isolation and none could be run there, which left the comparison with
its two lower branches (Section~\ref{sec:impl}).}
\Description{Diagram of a probe whose traffic crosses the wireless link
and then branches. One branch goes to a local reference. The other
crosses the building uplink and branches again, into a path to a cloud
reference and a path to a real service. A legend explains how the three
verdicts are read together.}
\label{fig:threeway}
\end{figure*}
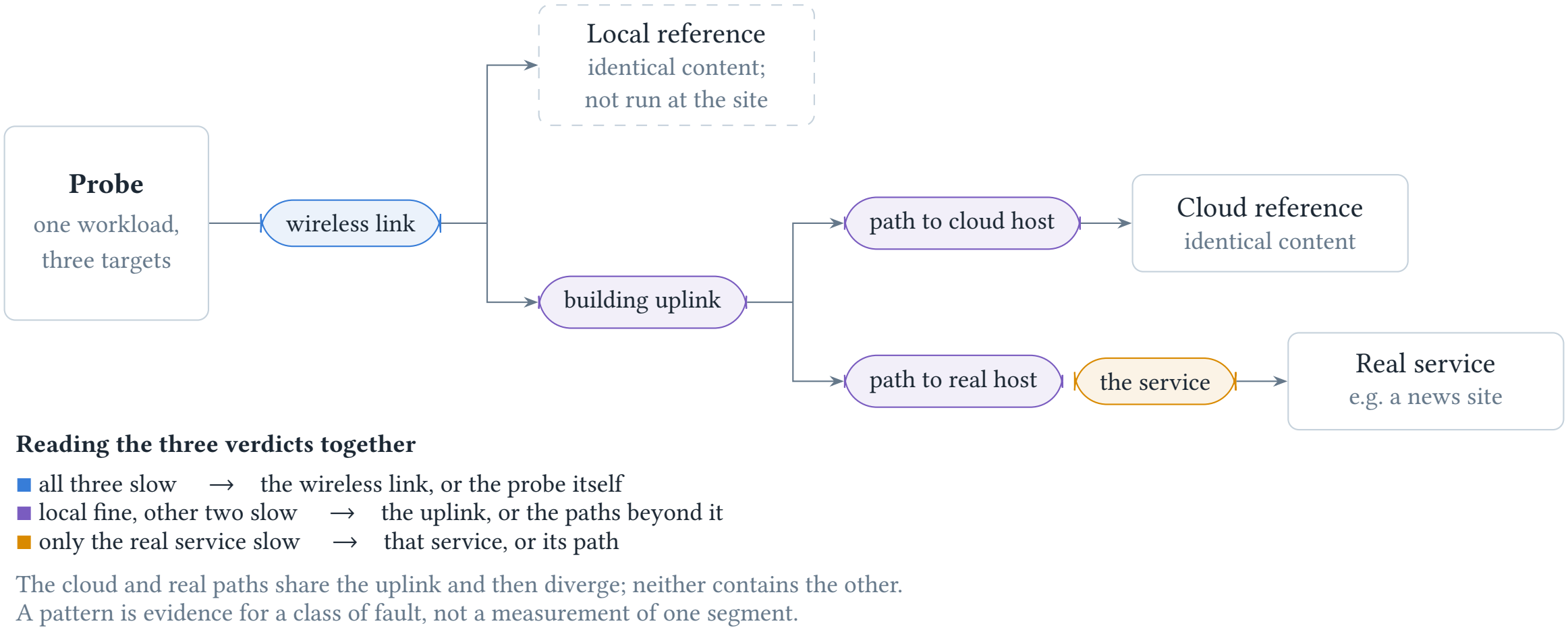

\subsection{The three-way endpoint split}
\label{sec:threeway}

Each time they are scheduled, the web, video and download workloads run
against three endpoints (Figure~\ref{fig:threeway}): a reference server
wired to the same local network, the same content on a cloud host under
the project's control, and a real third-party service. Email runs
against the real mail service only, because the reference server does
not serve IMAP. The local run crosses only the wireless link and the
local network. The cloud run crosses the wireless link, the building's
uplink and the path to the cloud host. The real run crosses the wireless
link, the uplink and a different path to a different host.

The attribution rule reads the three verdicts together. If all three are
slow, the cause is taken to be the wireless link or the probe itself. If
the local run is fine and the other two are slow, it is the uplink or
the paths beyond it. If only the real run is slow, it is that service or
its path. The rule rests on three assumptions. All three paths share the
wireless link. The cloud and real paths also share the uplink, then
diverge, and neither contains the other. And because the two references
serve identical content, a difference between them is mostly a
difference in the network, although hosting, server load, proxies and
connection behaviour can differ too. The pattern is evidence for a class
of fault; it does not identify a physical segment uniquely. The
comparison is between verdicts, not between times: the design never
subtracts one endpoint's time from another's and reads the remainder as
the time spent in a segment.

The three endpoints are deliberately not equivalent. The two references
serve identical content, while the real endpoint is a live news site of
several megabytes across dozens of requests. It was kept because the
project's claim is about what people experience, and it is judged
against its own thresholds. Section~\ref{sec:faultinjection} shows the
cost of that choice.

\subsection{Data contract}
\label{sec:contract}

Every measurement is one JSON record whose format is fixed by a JSON
Schema that the backend checks on arrival. A record carries a timestamp,
a probe identifier, a \texttt{run\_id} shared by every run of one
scheduled round, the workload, the endpoint, the measurements, the
probe's WiFi state (channel, signal strength, and the network's name and
access point address, SSID and BSSID) and a success flag with an error
message. Failed runs are stored, not dropped. The shared
\texttt{run\_id} is what makes the paired analysis of
Section~\ref{sec:deployment} possible: the five baseline probes of a
cycle can be compared with one another. A second kind of probe, an iOS
client built late in the project, posted records that the backend
accepted unchanged.

\subsection{Baseline destination classes}
\label{sec:baseline}

Beneath the application workloads, the baseline runs every ten seconds.
In each cycle it probes five destination classes in parallel, with ten
packets per class 0.2~s apart, all sharing one \texttt{run\_id}: the
default gateway, the two public anchors (1.1.1.1 and 8.8.8.8), a large
content delivery network (CDN, www.google.com) and the project's cloud
reference. The first four are probed with ICMP echo, the packets
\texttt{ping} sends. The cloud reference is probed by timing TCP
handshakes to port 443, because its host drops ICMP; each record names
its method, so the two are never mixed. Each class reports round-trip
time, jitter, packet loss and name-resolution (DNS) time.

The classes are chosen so that the pattern of failure across them is
informative. Packet loss or delay at the gateway alone points to the
local network. Loss to one distant destination points to that path or
provider. Loss to every destination except the gateway points to the
site's uplink. The gateway round trip stands in for the wireless link,
and it is a proxy, not a measurement of the radio: an echo to the
gateway crosses the probe's own network stack, the radio, the local
network and the router's handling of echo requests, which routers treat
as lower priority than forwarding and may rate-limit. Packet loss at the
gateway is therefore not a lower bound on loss through it, and its delay
includes time that forwarded traffic does not pay.
Section~\ref{sec:deployment} treats it as a proxy for the local path
throughout.

Two more probes run alongside the application workloads. The first tests
latency under load. It takes the median TCP-handshake round trip to a
public anchor while the link is idle and again during a download that
fills the link, and reports the added delay and the download rate
achieved. This is the effect Gettys and Nichols named
bufferbloat~\cite{gettys2012bufferbloat}, measured far more simply than
the IETF responsiveness draft specifies~\cite{ietf-responsiveness}: one
download and one anchor, not that draft's procedure. The second is a
path probe. Every five minutes it runs \texttt{mtr}, a traceroute tool,
in TCP mode to port 443 and records the round trip and packet loss at
each router (\emph{hop}) on the path, with lookups of each hop's name
and network (reverse DNS and AS number) that let the dashboard name its
operator and role.

A hop can report a larger round trip than it adds to forwarded traffic,
because a router may answer an expired packet slowly, limit how often it
answers, or reply by a different path. Govindan and Paxson found few
routers slow at generating these replies~\cite{govindan2002icmp}, but in
a single trace the effect is visible. Before the dashboard reports the
largest increase between consecutive hops as the likely bottleneck, it
therefore caps each hop's round trip at the smallest value seen at that
hop or any later one, a \emph{suffix minimum}: delay that is really on
the path must be paid by every later hop. This is a display heuristic,
not a validated correction. It can be wrong when hops reply by different
paths or are probed at different moments, so the result is reported as
the likeliest bottleneck and the raw trace is kept.

\subsection{Attribution}
\label{sec:attribution}

The dashboard's answer to \emph{where is the slowness} is computed over
the records in a time window by functions that take records in and give
a result out, with no database access, so that they can be tested on
exact inputs. The steps are these. For each workload and endpoint, the
median of the headline metric over the runs that completed is given a
verdict of good, degraded or poor against the thresholds in
Table~\ref{tab:thresholds}. If every attempt of a workload against an
endpoint failed, that endpoint is marked failed; if only some failed,
unstable. Each workload then casts at most one vote for a segment
(Figure~\ref{fig:threeway}): the wireless link if its local endpoint is
degraded, poor, failed or unstable, otherwise the internet path if its
cloud endpoint is, otherwise the third party if its real endpoint is.
The segment with the most votes is reported as the likely cause. A tie
goes to the segment nearest the user, because a fault there would also
explain the symptoms seen beyond it. Each segment also shows a state of
its own: \emph{no data}, not a misleading \emph{OK}, when its endpoint
produced no record in the window; \emph{down} when every attempt of some
workload against it failed; and \emph{unstable} when some did. With no
local endpoint, the wireless link is judged by the median round trip to
the gateway against its line in Table~\ref{tab:thresholds}
(Section~\ref{sec:impl}). The thresholds were engineering estimates set
before any sustained data existed. The real endpoint's web and video
thresholds are higher than the references' because its content is
heavier.

\begin{table}[t]
\caption{Good and poor thresholds. Values between the two classify as
degraded. The real site's web and video lines are higher than the
reference's because it serves far heavier content. The 2,000~ms good
line for the reference page is the preferred target ITU-T G.1010 gives
for a web page \cite{itu-g1010}.}
\label{tab:thresholds}
\footnotesize
\begin{tabular}{@{}lrr@{}}
\toprule
\textbf{Measure} & \textbf{Good} & \textbf{Poor} \\
\midrule
Web page load, reference page (ms) & $\le 2{,}000$ & $> 5{,}000$ \\
Web page load, real site (ms)      & $\le 4{,}000$ & $> 8{,}000$ \\
Video startup, reference (ms)      & $\le 3{,}000$ & $> 8{,}000$ \\
Video startup, real site (ms)      & $\le 4{,}000$ & $> 10{,}000$ \\
Email fetch (ms)                   & $\le 1{,}000$ & $> 3{,}000$ \\
Download throughput (Mbit/s)       & $\ge 20$      & $< 5$ \\
Gateway round trip, median (ms)    & $\le 10$      & $> 30$ \\
Added latency under load (ms)      & $\le 30$      & not defined \\
\bottomrule
\end{tabular}
\end{table}

\subsection{Health score}
\label{sec:health}

A resident or an operator who asks whether the network is acceptable
wants one number. The health score is an engineering index built for
that purpose. It is not a measured quality of experience. Each
workload's median headline metric is mapped to a quality index from 0 to
100, tied to the five-point mean opinion score (MOS) scale of ITU-T
P.800~\cite{itu-p800}: the good threshold maps to MOS~4, a score of 75,
and the poor threshold to MOS~2, a score of 25, with a straight line
between them and a flat value beyond. This reuses the thresholds of
Table~\ref{tab:thresholds} instead of adding a second set of judgements.
A fifth component, responsiveness, is derived from added latency under
load and packet loss. The five are combined in a weighted mean: video
0.35, because rebuffering has the largest measured effect on engagement
of any quality metric~\cite{dobrian2011video}; responsiveness 0.25,
because latency under load is what interactive use feels and no
page-load figure captures it~\cite{sundaresan2011broadband}; web 0.25,
because waiting time drives web experience~\cite{itu-g1030}; download
0.10, because G.1010 gives bulk transfer a preferred target of 15~s
against 2~s for a web page~\cite{itu-g1010}; and email 0.05, because a
user rarely waits on a fetch (G.1010 gives mail server access the same
targets as web browsing). A component with no data is left out and the
remaining weights are rescaled. The weights are design choices tied to
those sources, not fitted values. The MOS scale anchors the index and
does not validate it, and no test with people was carried out. The
dashboard's labels, from \emph{excellent} to \emph{bad}, are bands of
the index and as heuristic as the index.

Every figure above is a median, and a median describes only the runs
that produced a number. The weighted mean is therefore reported as
\emph{quality}: how good the service was when it worked. The published
score is quality multiplied by \emph{availability}, the share of web,
video, email and download runs in the window that completed.
Availability is a completion rate over recorded attempts, not a share of
elapsed time: a period in which the probe recorded nothing does not
lower it, and a workload that runs more often counts for more. Baseline
probes are left out of it because they complete even when they report
total packet loss, which is what makes them useful during an outage.
Section~\ref{sec:outage} describes the incident that forced this
separation.

\subsection{Dashboard}
\label{sec:dashboard}

Three views read the same summary from the backend. An overview page
carries the score, a one-line verdict and a traffic light per segment. A
diagnostic page shows each workload's result at each endpoint, the
evidence behind the attribution, the path and the latency-under-load
result. Its window is one hour, 24 hours or seven days and may end at
any past instant, so the stored records of a past period can be run
through the current code again; this paper calls that a \emph{replay}. A
Grafana board plots every metric with one line per endpoint.

\section{Implementation}
\label{sec:impl}

\paragraph{Probe.}
The probe is a Raspberry~Pi~4 running Raspberry Pi OS Lite. Its software
is a Python~3.11 program of about 1,300 lines, with 1,700 lines of
tests, that \texttt{systemd} starts on boot and restarts after a crash.
The web workload loads each page in the Chromium browser, run without a
display and driven by the Playwright library, and reports the phases the
browser itself times (name resolution, connection, first byte, page
load) through its Navigation Timing
interface~\cite{w3c-navigation-timing}, following the argument of Wang
et al.\ that only a browser engine sees what the user waits
for~\cite{wang2013wprof}. Each load starts a fresh browser process, so
nothing is cached from one load to the next, and waits for the page's
load event for up to 60~s. The endpoints of a workload are visited one
after another, the reference before the real service.

The video workload finds the stream's address, with \texttt{yt-dlp}
where needed, reads its bitrate with \texttt{ffprobe}, and downloads it
while simulating a player that starts after two seconds of video have
arrived and then plays at normal speed. A stall, or \emph{rebuffer}, is
recorded whenever the simulated player runs out of downloaded video. No
rebuffer occurred in the month. The two clips are of about 0.8~Mbit/s
and were delivered at a median of 33 to 34~Mbit/s
(Section~\ref{sec:deployment}), so the delivery rate would have had to
fall by about 98\% before the simulated player stalled. The workload
therefore also reports the delivery rate, its ratio to the clip's
bitrate (a median of 43 to 45 once it was recorded) and the highest
standard resolution that rate would sustain. The page's load event and
the simulated fixed-bitrate player are stand-ins: they do not measure
when a page became usable or how a player that adapts its bitrate would
have behaved.

The email workload times an IMAP fetch from a mailbox created for the
purpose. The download workload times the transfer of a file of known
size, a 50~MB public test file at the real endpoint. The baseline runs
\texttt{ping} asking for ten replies at 0.2~s spacing, with a 2~s wait
for each reply and a 6~s limit on the whole run, and for the cloud class
it times TCP connections. The path probe runs \texttt{mtr}. The baseline
runs every 10~s, the web, video and email workloads and the path probe
every 5~min, and the download and latency-under-load tests every hour:
at five-minute intervals those two would have moved 15 to 20~GB a day
over a shared uplink and onto a public test server, against about
3.5~GB a day as deployed. Every record is first written to a local
SQLite database and uploaded by a separate process that retries with
growing delays, so records survive a loss of connectivity; this is the
probe's \emph{store-and-forward buffer}.

\paragraph{Backend and dashboards.}
The backend is about 950 lines of Python (FastAPI) on Azure App Service,
489 of them the attribution and scoring module. It accepts a record only
with the probe's access token and only if the record matches the schema,
and stores it in PostgreSQL with the TimescaleDB extension for time
series. Every Azure resource is declared in one Bicep file, so one
command creates the whole deployment and one deletes it. The student
subscription could not create resources in UK regions, so the backend
and the cloud reference run in Norway. In the paired period
(Section~\ref{sec:deployment}) the cloud reference's TCP handshake took
a median of 34.63~ms against 11.62~ms for ICMP to 1.1.1.1, so much of
the delay to the cloud reference is distance. The attribution rule
tolerates this because it compares results with thresholds, not with an
expected value. Every defect in Sections~\ref{sec:outage}
and~\ref{sec:faultinjection} was first reproduced as an automated test
on the exact records that exposed it, then fixed; there are 58 tests.

\paragraph{Reference content.}
The local server and the cloud endpoint run the same nginx web server
image. It serves a static page, a 60-second clip generated with
\texttt{ffmpeg} and a 25~MiB file of pseudo-random bytes, all produced
when the image is built, so the content needs no licence and is
byte-identical wherever it runs.

\paragraph{What could not run at the site, and what it cost.}
The deployment network enforces \emph{client isolation}: two devices on
it cannot reach each other. The room's router also passed the site
network through to its wired ports instead of creating a private network
of its own. No local reference server could therefore be run, and the
three-way comparison of Figure~\ref{fig:threeway} had two endpoints for
the whole deployment. The cost is material. In this paper the wireless
link is judged by the round trip and packet loss of the gateway probe, a
proxy for the local path (Section~\ref{sec:baseline}). There was no
like-for-like throughput measurement and no application-level
measurement of the link alone. The deployed rule took its link verdict
from the first hop of the traceroute instead, a substitution that the
fault injection of Section~\ref{sec:faultinjection} later showed to be
unsafe.

The site also puts a sign-in page, a \emph{captive portal}, in front of
internet access, and its session expired about once a day. A small
module opens the portal in the installed browser every two minutes,
accepts its terms and confirms that the internet is reachable again.
Without it the unattended run would have ended within a day.

One interruption was the probe's own. On 17 August 2026 its SD card was
corrupted during a software update and the program failed as it started.
\texttt{systemd} restarted it every ten seconds and it failed the same
way each time, for 19.88~h, until this was noticed by
hand.\footnote{The dissertation describes this outage as approximately
24 hours; the stored records put it at 19~h~52~min.} Restarting after a
crash does not help when every start crashes, and nothing watched for
the probe going silent.

\section{Deployment results}
\label{sec:deployment}

The probe ran as an ordinary client on the resident WiFi network of a
student accommodation building in Liverpool. The period analysed is the whole
unattended run at that site, from the first baseline cycle at 10:54 UTC on
28 July 2026 to the last at 20:54 UTC on 28 August 2026, when the probe was
restarted for the fault injection session. 795{,}252 baseline records
carry that site's label; records from other sites are excluded.\footnote{The
dissertation and the repository README describe the same deployment as
29 days with 706{,}265 records across all workloads, because their
database snapshot was taken at 16:00 UTC on 25 August 2026 while the
probe was still running. This paper uses the complete run to 28 August;
the counts here are baseline records only. Comparisons with the
dissertation's Tables~4.6 and~4.7 use the rows of its own window.}

Every ten seconds the baseline probes five destination classes: the
default gateway, the two public anchors (1.1.1.1 and 8.8.8.8), a large
CDN (www.google.com) and the project's cloud reference in Oslo, the last
by TCP handshake. Each probe reports a round-trip time (RTT) averaged
over its packets. The five probes of a cycle share one run identifier,
start within a few milliseconds of each other and finish within about
2.5~s, so the classes can be compared cycle by cycle (\emph{paired}),
not only as five separate distributions over the whole period
(\emph{pooled}). Until 10:28 UTC on 18 August 2026 the baseline probed
only the two anchors, one after the other, with five packets at
one-second spacing. From that cycle on it probed all five classes in
parallel with ten packets at 0.2~s spacing. The paired analysis
therefore covers the 89{,}455 complete five-class cycles from that
instant to the end of the run, 10.4 days, called the \emph{paired
period}. The 173{,}985 two-anchor cycles before it are used only in the
coverage and packet loss figures. 89{,}357 of the complete cycles have a
valid RTT in all five classes; the rest fall almost entirely in the loss
of off-site reachability of 18 to 19 August 2026, which
Section~\ref{sec:outage} describes and which is called the uplink
failure from here on.

\subsection{Gateway and far-end round trips in the same cycle}

\begin{figure*}[t]
\centering
\includegraphics[width=\textwidth]{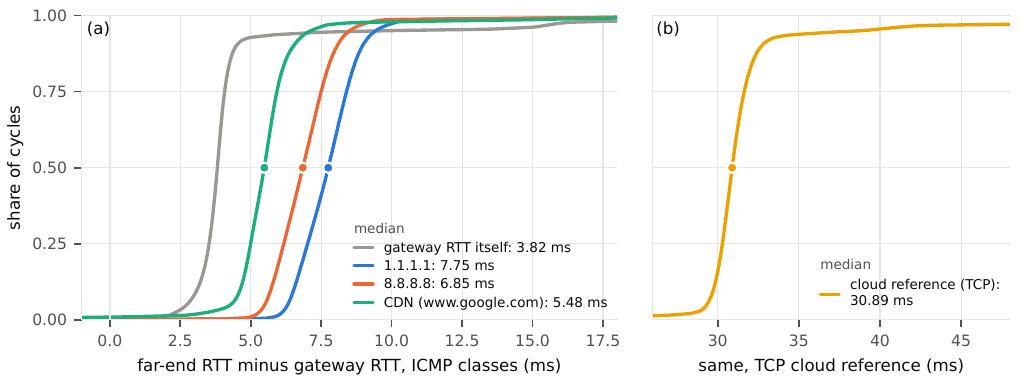}
\caption{Empirical distribution of the within-cycle difference, far-end
RTT minus default-gateway RTT in the same ten-second cycle, over 89{,}358
cycles from 18 to 28 August 2026. (a) The three ICMP classes, with the
gateway RTT itself in grey for scale. (b) The TCP handshake to the cloud
reference. The median of each series is marked and given in the legend.}
\Description{Two empirical cumulative distribution plots. Panel (a) shows
far-end minus gateway round-trip time for the two public anchors and the
CDN, with the gateway RTT in grey; panel (b) shows the same difference for
the TCP cloud reference.}
\label{fig:paired}
\end{figure*}

\begin{table*}[t]
\caption{Round-trip time in the paired period, 18 August 2026 10:28 UTC to
28 August 2026 20:54 UTC, in milliseconds. Top: the within-cycle
difference, far-end RTT minus default-gateway RTT measured in the same
ten-second cycle, one value per cycle and class; \emph{negative} is the
share of cycles in which the far end answered faster than the gateway.
Bottom: pooled quantiles of the same cycles, with the dissertation's
Table~4.6 alongside. That table pooled its whole window, which ended at
16:00 UTC on 25 August 2026, so for the two public anchors three quarters
of its samples come from before 18 August 2026, when the anchor was probed
with five packets at one-second spacing.}
\label{tab:paired}
\small
\begin{tabular}{lrrrrrrr}
\toprule
\textbf{Class minus gateway} & \textbf{Cycles} & \textbf{p5} & \textbf{p25} & \textbf{p50} & \textbf{p75} & \textbf{p95} & \textbf{Negative} \\
\midrule
Public anchor 1.1.1.1 (ICMP)          & 89{,}358 &  6.31 &  7.02 &  7.75 &  8.40 &  9.43 & 0.12\% \\
Public anchor 8.8.8.8 (ICMP)          & 89{,}358 &  5.50 &  6.17 &  6.85 &  7.50 &  8.51 & 0.13\% \\
Large CDN, www.google.com (ICMP)      & 89{,}357 &  4.29 &  5.00 &  5.48 &  5.93 &  7.12 & 0.89\% \\
Cloud reference, Oslo (TCP handshake) & 89{,}362 & 29.33 & 30.29 & 30.89 & 31.63 & 38.70 & 0.09\% \\
\bottomrule
\end{tabular}

\medskip

\begin{tabular}{lrrrrrrrrr}
\toprule
 & & \multicolumn{4}{c}{\textbf{Paired period, pooled}} & \multicolumn{4}{c}{\textbf{Dissertation Table 4.6}} \\
\cmidrule(lr){3-6} \cmidrule(lr){7-10}
\textbf{Class} & \textbf{Cycles} & \textbf{p5} & \textbf{p50} & \textbf{p95} & \textbf{p99} & \textbf{p50} & \textbf{p95} & \textbf{p99} & \textbf{Samples} \\
\midrule
Default gateway (ICMP)                & 89{,}455 &  2.74 &  3.82 &  9.68 &  27.83 &  3.83 &  6.61 &  23.67 &  61{,}810 \\
Public anchor 1.1.1.1 (ICMP)          & 89{,}358 & 10.30 & 11.62 & 19.94 &  39.97 & 13.12 & 14.12 &  39.38 & 235{,}694 \\
Public anchor 8.8.8.8 (ICMP)          & 89{,}358 &  9.49 & 10.70 & 19.03 &  38.58 & 11.96 & 12.53 &  37.91 & 235{,}692 \\
Large CDN, www.google.com (ICMP)      & 89{,}357 &  8.64 &  9.38 & 18.38 &  37.93 &  9.38 & 16.82 &  33.91 &  61{,}718 \\
Cloud reference, Oslo (TCP handshake) & 89{,}362 & 33.41 & 34.63 & 48.47 & 138.07 & 34.68 & 45.48 & 137.29 &  61{,}718 \\
\bottomrule
\end{tabular}
\end{table*}

Table~\ref{tab:paired} and Figure~\ref{fig:paired} give the
\emph{within-cycle difference}: the RTT to a distant destination (the
far end) minus the RTT to the gateway in the same cycle. For 1.1.1.1 it
has a median of 7.75~ms over 89{,}358 cycles and a 5th to 95th
percentile range of 6.31 to 9.43~ms (95\% interval on the median 7.73
to 7.77~ms, from the bootstrap described below); the median is 6.85~ms for 8.8.8.8, 5.48~ms
for the CDN and 30.89~ms for the TCP handshake to the cloud reference. The
spread of the difference is much smaller than the spread of either RTT:
its 5th to 95th percentile width is 3.1~ms for 1.1.1.1, against 6.9~ms for
the gateway RTT and 9.6~ms for the anchor RTT. The percentiles are of
per-run averages over ten packets, not of individual packets.

The slow cycles of the two coincide. Call an RTT \emph{high} when it is
above its own 95th percentile over the paired cycles (9.69~ms for the
gateway, 19.94~ms for 1.1.1.1). The anchor RTT is high in 92.2\% of the
cycles in which the gateway RTT is high (95\% interval 91.2 to 92.9\%)
and in 0.41\% of the other cycles; by definition it is high in 5\% of
all cycles.
8.8.8.8 behaves the same (92.4\%), the CDN (82.3\%) and the cloud
reference (77.5\%) less strongly. When the gateway is high, the median 1.1.1.1 RTT
is 23.82~ms against 11.56~ms otherwise, while the median difference is
unchanged, 7.70 against 7.75~ms: the anchor RTT rises by about as much as
the gateway RTT. The CDN rises by less (difference 3.76 against 5.50~ms)
and the TCP cloud reference by more (38.30 against 30.87~ms, the pattern
a retransmitted handshake would give). The rank correlation over all
cycles is low (Spearman 0.12 for 1.1.1.1, 0.26 on cycle-to-cycle changes)
because in most cycles both RTTs move by fractions of a millisecond and
those movements are unrelated. The two are related in their slow cycles,
and slow spells are short: the 4{,}468 cycles with a high gateway RTT
form 4{,}043 runs of consecutive cycles, of median length one cycle
(maximum five). The probe's own traffic is associated with a small part
of them. The times at which its hourly transfer tests and its two
busiest application workloads ran were estimated from the schedule, not
from the workloads' own timestamps. Cycles in those times have a high
gateway RTT in 6.1\% and 7.6\% of cases, against 4.3\% elsewhere, with
the same median difference in all three groups.

The intervals above come from a bootstrap that resamples whole UTC
days, 11 days and 1{,}000 resamples, with both thresholds recomputed in
each resample. They describe the uncertainty from which days this
deployment happened to observe, and say nothing about other networks.
The 92.2\% figure depends on the threshold: with each series' own 90th
percentile it is 72.3\%, with the 99th 84.5\%. Day by day, with the
95th-percentile thresholds of the whole period, it ranges from 87.1\% to
94.4\% and the median difference from 7.69 to 7.80~ms. Leaving any one
day out and recomputing the thresholds moves the overall figure between
91.9\% and 92.4\%.

These are statements about two measurements moving together within a
cycle; they do not show what causes either. The gateway RTT is a proxy
for the local path (Section~\ref{sec:baseline}), not a measurement of
the wireless link alone, and no difference here is read as the time
spent inside any one segment.

\paragraph{What changed against the dissertation.}
The dissertation reported the gateway median of 3.83~ms as 29\% of a
13.12~ms median path to 1.1.1.1. Those two medians came from different
periods and probe settings: the gateway's from the 61{,}810 samples
after 18 August 2026, the anchor's from 235{,}694 samples over the whole
run, three quarters of them from before 18 August 2026, when the anchor's
median was 13.33~ms. In the same cycles the
medians are 3.82 and 11.62~ms, the difference between the two overall
medians is 7.81~ms and the median within-cycle difference 7.75~ms, where the
dissertation's subtraction gave 9.29~ms. The same-period ratio of medians
is 33\% (within-cycle median 0.33), but a ratio of two round trips that
share the wireless hop is not a share of anything, and this paper reports
the differences. The dissertation also took the ratio of the 95th
percentile to the median, 1.75 for the gateway against 1.08 for the
anchor, to mean that the wireless link carried nearly all the
variability. In the paired period the ratios are 2.54 and 1.72, so the
anchor's slow cycles are not rare, and what the paired data show is that
the two are slow together.

\subsection{Coverage, kept apart from task success}

\begin{table*}[t]
\caption{Coverage, kept apart from task success. Scheduled cycles are the
ten-second slots between the first cycle at the site, 28 July 2026
10:54 UTC, and the last, 28 August 2026 20:54 UTC (the cadence between
produced cycles has a median of 10.00~s and a 99th percentile of 10.01~s);
coverage is cycles produced over cycles scheduled. Failed and no-reply are counts of baseline
records on that day, and completed is the share of the day's records that
completed; a no-reply record is a completed measurement of an unreachable
destination. The rows are the whole window, the four days below 99.5\% and
the second day of the uplink failure; each of the other 27 days is at
99.54\% or above. Below: every gap longer than 60~s between consecutive
cycles.}
\label{tab:coverage}
\small
\begin{tabular}{lrrrrrr}
\toprule
\textbf{Period} & \textbf{Scheduled} & \textbf{Produced} & \textbf{Coverage} & \textbf{Failed} & \textbf{No reply} & \textbf{Completed} \\
\midrule
Whole window, 28 July to 28 August 2026 & 271{,}436 & 263{,}443 & 97.06\% & & & \\
\midrule
11 August 2026                          &   8{,}640 &   8{,}584 & 99.35\% &   0 &  24 & 100.00\% \\
17 August 2026                          &   8{,}640 &   5{,}256 & 60.83\% &   0 &   0 & 100.00\% \\
18 August 2026                          &   8{,}640 &   4{,}711 & 54.53\% & 158 & 182 &  99.33\% \\
19 August 2026                          &   8{,}640 &   8{,}627 & 99.85\% &  14 &  14 &  99.97\% \\
21 August 2026                          &   8{,}640 &   8{,}159 & 94.43\% &   0 &   1 & 100.00\% \\
\bottomrule
\end{tabular}

\medskip

\begin{tabular}{lllr}
\toprule
\textbf{Last cycle before the gap (UTC)} & \textbf{First cycle after it (UTC)} & \textbf{What the probe was doing} & \textbf{Gap} \\
\midrule
28 July 2026 11:22:38   & 28 July 2026 11:23:58   & first hour of the run                            & 80~s \\
17 August 2026 14:35:48 & 18 August 2026 10:28:46 & probe outage; ends at the first five-class cycle & 19.88~h \\
21 August 2026 10:13:21 & 21 August 2026 10:14:58 & moved to a laptop hotspot for an experiment      & 97~s \\
21 August 2026 10:16:18 & 21 August 2026 11:28:41 & on the hotspot                                   & 1.21~h \\
21 August 2026 11:28:51 & 21 August 2026 11:33:18 & moved back                                       & 267~s \\
21 August 2026 11:33:18 & 21 August 2026 11:34:31 & moved back                                       & 73~s \\
\bottomrule
\end{tabular}
\end{table*}

\begin{table*}[t]
\caption{Packet loss as lost over sent in the paired period, after
18 August 2026 10:28 UTC. The probe stored ping's loss percentage, not its
counters, so sent and lost are estimates under the probe's settings (ten
replies requested at 0.2~s spacing, a 6~s deadline): a run that reached ten
replies is taken to have sent ten packets plus one per lost packet, which is
what the reported percentages imply (Section~\ref{sec:deployment}); a run whose
percentage would need more than 30 packets was ended by the deadline and is
counted as 30 sent (two runs per class); 4 to 6 runs per class
report a percentage consistent only with more than ten replies and take the
smallest consistent count. The remaining ambiguity moves no rate by more
than a few thousandths of a percentage point. A run with no reply at all is
kept apart; \emph{incl.\ no reply} counts each such run as 30 packets sent
and lost, the most the deadline allows (counting ten, the dissertation's
assumption, gives 0.39\% for the anchors). Failed runs, in which DNS
resolution did not succeed, sent no packets. The cloud reference sends ten
TCP handshakes, and a lost handshake is one the operating system did not
complete within 2~s, so its row is not comparable with ICMP loss. The last
two columns use only the rows of the dissertation's window, which ended at
16:00 UTC on 25 August 2026: lost over sent on those rows, and the
dissertation's Table~4.7, which averaged per-run percentages with every
no-reply run counted as 100\%.}
\label{tab:loss}
\small
\begin{tabular}{lrrrrrrrrr}
\toprule
 & \multicolumn{7}{c}{\textbf{Paired period}} & \multicolumn{2}{c}{\textbf{Dissertation window}} \\
\cmidrule(lr){2-8} \cmidrule(lr){9-10}
\textbf{Class} & \textbf{Runs} & \textbf{No reply} & \textbf{Failed} & \textbf{Sent} & \textbf{Lost} & \textbf{Lost/sent} & \textbf{Incl.\ no reply} & \textbf{Lost/sent} & \textbf{Table 4.7} \\
\midrule
Default gateway (ICMP)                & 89{,}455 &  0 &  0 & 897{,}233 & 2{,}685 & 0.299\% & 0.299\% & 0.305\% & 0.270\% \\
Public anchor 1.1.1.1 (ICMP)          & 89{,}456 & 97 &  0 & 896{,}082 & 2{,}494 & 0.278\% & 0.601\% & 0.332\% & 0.380\% \\
Public anchor 8.8.8.8 (ICMP)          & 89{,}456 & 97 &  0 & 896{,}071 & 2{,}482 & 0.277\% & 0.600\% & 0.287\% & 0.324\% \\
Large CDN, www.google.com (ICMP)      & 89{,}375 & 17 & 81 & 895{,}680 & 2{,}105 & 0.235\% & 0.292\% & 0.235\% & 0.229\% \\
Cloud reference, Oslo (TCP handshake) & 89{,}363 &  0 & 93 & 893{,}630 &      49 & 0.005\% & 0.005\% & 0.007\% & 0.007\% \\
\bottomrule
\end{tabular}
\end{table*}

\emph{Coverage} is the number of cycles the probe produced over the
number scheduled, where a scheduled cycle is a ten-second slot between
the first cycle at the site and the last. \emph{Task success} is the
share of produced records that completed. A record completes either with
replies or with \emph{no reply}, when the probe ran and nothing
answered. A record \emph{failed} when the probe could not run at all,
in practice because the destination's name did not resolve. Coverage and
task success are kept apart for two reasons. A record that completed
with no reply is a successful measurement of an unreachable destination.
And the probe can lose cycles while it keeps running: a run with no
reply waits for \texttt{ping}'s 6~s limit and a destination given by
name waits for name resolution, so the cycles that overlap are skipped.
Table~\ref{tab:coverage} gives the whole period, the days below 99.5\%
and the gaps.

Over the whole period 263{,}443 cycles were produced of 271{,}436
scheduled, a coverage of 97.06\%. Most of the shortfall is one probe
outage, from the last cycle at 14:35:48 UTC on 17 August 2026 to the first
at 10:28:46 UTC on 18 August 2026: 19.88~h (19~h 52~min). Its cause was
the crash on every start described in Section~\ref{sec:impl}. The four gaps on
21 August 2026, the longest 1.21~h, are the probe being moved to a laptop
hotspot and back for an experiment. Cycles were also lost while the
probe kept running. On 11 August 2026 the probe is 56 cycles short with no gap over
60~s: in 13 cycles between 11:19 and 11:29 UTC the anchors got no reply,
24 no-reply records in all. The gateway class did not exist yet, so a
scan for the pattern of Section~\ref{sec:outage} could not see this
episode. During the uplink failure, from 23:23 UTC on 18 August 2026
to 00:03 UTC on 19 August 2026, in the 92 cycles in which both anchors
had no reply, consecutive cycles were a median 30~s apart instead of
10~s.

Task success tells a different story. On 17 August 2026 every produced record
completed; the day is short of cycles, not of successful measurements. On
18 August 2026, with coverage at 54.53\%, 158 of 23{,}555 records failed
and 182 had no reply, and 99.33\% completed; on 19 August 2026, 14 failed,
14 had no reply and 99.97\% completed. The failed runs are DNS failures
for the CDN and cloud hostnames. The no-reply runs are the uplink
failure, in which both anchors got no reply in 92 cycles while the CDN
and cloud hostnames failed at resolution, the 11 August 2026 episode, and
five cycles at 11:34 UTC on
28 August 2026 in which every ICMP destination beyond the gateway was
silent while the TCP cloud reference answered in all five: ICMP dropped
somewhere beyond the gateway, not a loss of connectivity, and one more
reason to keep the two probe methods apart.

\subsection{Packet loss as lost over sent}

The dissertation reported packet loss as the mean of the loss
percentages of individual runs (its Table~4.7), which gives a five-packet run before 18 August 2026 the
same weight as a ten-packet run after it and counts a run with no reply
as 100\%. Table~\ref{tab:loss} instead estimates lost and sent
packets per run from ping's printed loss percentage under the probe's
settings, keeps the runs with no reply in their own column because their
sent count is unknown, and adds a second figure that counts each of
them as thirty packets sent and lost, the most a 6~s deadline allows. The
probe did not store ping's counters, so the totals are estimates whose
assumptions the caption states; the alternative readings of the few
ambiguous runs move no rate by more than a few thousandths of a
percentage point. The rule for sent packets follows what the runs
reported, not a reading of \texttt{ping}'s documentation. Before
18 August 2026, five-packet runs at one-second spacing reported only 0,
20, 40 or 100\%, which is five packets sent. From then on, a ten-packet
run at 0.2~s spacing that lost anything most often reported 9.09\%, one
in eleven, so a lost reply was followed by an extra packet. Why the two
configurations behaved differently was not established, and the probe
recorded neither \texttt{ping}'s counters nor its version.
The cloud reference's ten TCP handshakes are not comparable with ICMP
loss, since a lost handshake is one the operating system did not complete
within 2~s.

With runs weighted by packets sent and the no-reply runs of the uplink
failure kept apart, loss while the destination was reachable is 0.30\% at
the gateway, 0.28\% at each anchor, 0.235\% at the CDN and 0.005\% at the
cloud reference. The gateway is the highest of the ICMP classes. The
dissertation's reading, that loss at the gateway is the least any other
destination can show, does not hold in the paired data, and its own Table~4.7 already had
the CDN below the gateway: a router answering echo requests addressed to
itself can drop or rate-limit them, so echo loss at the gateway is not a
lower bound for loss on paths through it. The per-run average reproduces
the dissertation's values exactly on the rows of its own window (0.270\% at
the gateway, 0.380\% and 0.324\% at the anchors, 0.229\% at the CDN), so
the differences come from the method, not from the data.

\subsection{Application workloads}

The application records were exported after the baseline reanalysis:
55{,}012 records of the site over the run, read from the same database
table. Cut at the time of the dissertation's snapshot, the export
reproduces its Table~4.5 exactly, 49{,}238 records and 79 failed runs.\footnote{Computed
by \texttt{notes/application\_results.py}, which writes its report to
\texttt{data/processed/application-results.md}.}
Table~\ref{tab:application} gives, per workload and endpoint, the
records, the failed runs and the median and 95th percentile of the
headline metric. Over the run 80 runs failed, 0.15\%: 43 of them in the
uplink failure of 18 to 19 August 2026, 12 in the ten minutes on
11 August 2026 in which the anchors got no reply (described above), and
the remaining 25 spread over the month, 16 mailbox fetches that timed
out, eight video runs, five of them on 16 August 2026, and one page load
that exceeded its 60~s limit. Failures cluster where reachability was
lost; none is a sustained degradation of one service.

The two endpoints differ in content as much as in path. The cloud
reference page loaded in a median 237~ms against 3{,}330~ms for the real
news site, fourteen times longer for a page that is megabytes across
dozens of requests, over the same link and uplink; the 95th percentiles
are 609 and 4{,}066~ms. This difference in the size of the content is
why each endpoint has its own thresholds (Section~\ref{sec:threeway}). Video started in a
median 287~ms from the cloud reference and 126~ms from the real host, two
different clips on two different hosts rather than a like-for-like pair,
and no run of either rebuffered; the delivery rate, recorded from
18 August 2026, was a median 34.0 and 33.1~Mbit/s against clip bitrates
of about 0.8~Mbit/s, more than forty times what playback needed.
Download throughput was a median 56.9~Mbit/s from the cloud reference and
60.4~Mbit/s from the public test file. The
latency-under-load test added a median 29.7~ms to the idle round trip, 38.4~ms at the
95th percentile and 57.9~ms at most. 

\begin{table}[t]
\caption{Application workloads over the full unattended run, 28 July to
28 August 2026: records, failed runs in brackets, and the median and 95th
percentile of the headline metric over completed runs. A failed run stored
an error and no measurement. Web and video ran against the cloud reference
and the real service, email and the latency-under-load test against the
real service only, download against both; no local reference was available
at the site. The video's delivery rate was recorded from 18 August 2026
10:29 UTC, so its two rows cover 2{,}982 and 2{,}983 runs; no video run
rebuffered. Cut at the dissertation's snapshot of 25 August 2026, the same
export reproduces its Table~4.5 exactly: 49{,}238 records and 79 failed
runs.}
\label{tab:application}
\footnotesize
\setlength{\tabcolsep}{3pt}
\begin{tabular}{@{}>{\raggedright\arraybackslash}p{2.05cm}>{\raggedright\arraybackslash}p{2.25cm}rrr@{}}
\toprule
\textbf{Workload, endpoint} & \textbf{Metric} & \textbf{Records (failed)} & \textbf{p50} & \textbf{p95} \\
\midrule
Web, cloud               & page load, ms          & 8{,}799 (11) &    237 &    609 \\
Web, real                & page load, ms          & 8{,}799 (10) & 3{,}330 & 4{,}066 \\
Video, cloud             & startup, ms            & 8{,}798 (12) &    287 &    368 \\
                         & delivery rate, Mbit/s  &              &   34.0 &   36.1 \\
Video, real              & startup, ms            & 8{,}798 (16) &    126 &    246 \\
                         & delivery rate, Mbit/s  &              &   33.1 &   36.9 \\
Email, real              & fetch, ms              & 8{,}798 (25) &    592 & 1{,}366 \\
Download, cloud          & throughput, Mbit/s     &    742 (2)   &   56.9 &   60.8 \\
Download, real           & throughput, Mbit/s     &    741 (2)   &   60.4 &   64.1 \\
Latency under load, real & added latency, ms      &    740 (2)   &   29.7 &   38.4 \\
Path, real               & RTT to 1.1.1.1, ms     & 8{,}797 (0)  &   13.8 &   46.5 \\
                         & first-hop RTT, ms      &              &    4.9 &   20.4 \\
\midrule
All application runs     &                        & 55{,}012 (80) & \multicolumn{2}{r}{failure rate 0.15\%} \\
\bottomrule
\end{tabular}
\end{table}

\section{An unplanned loss of off-site reachability}
\label{sec:outage}

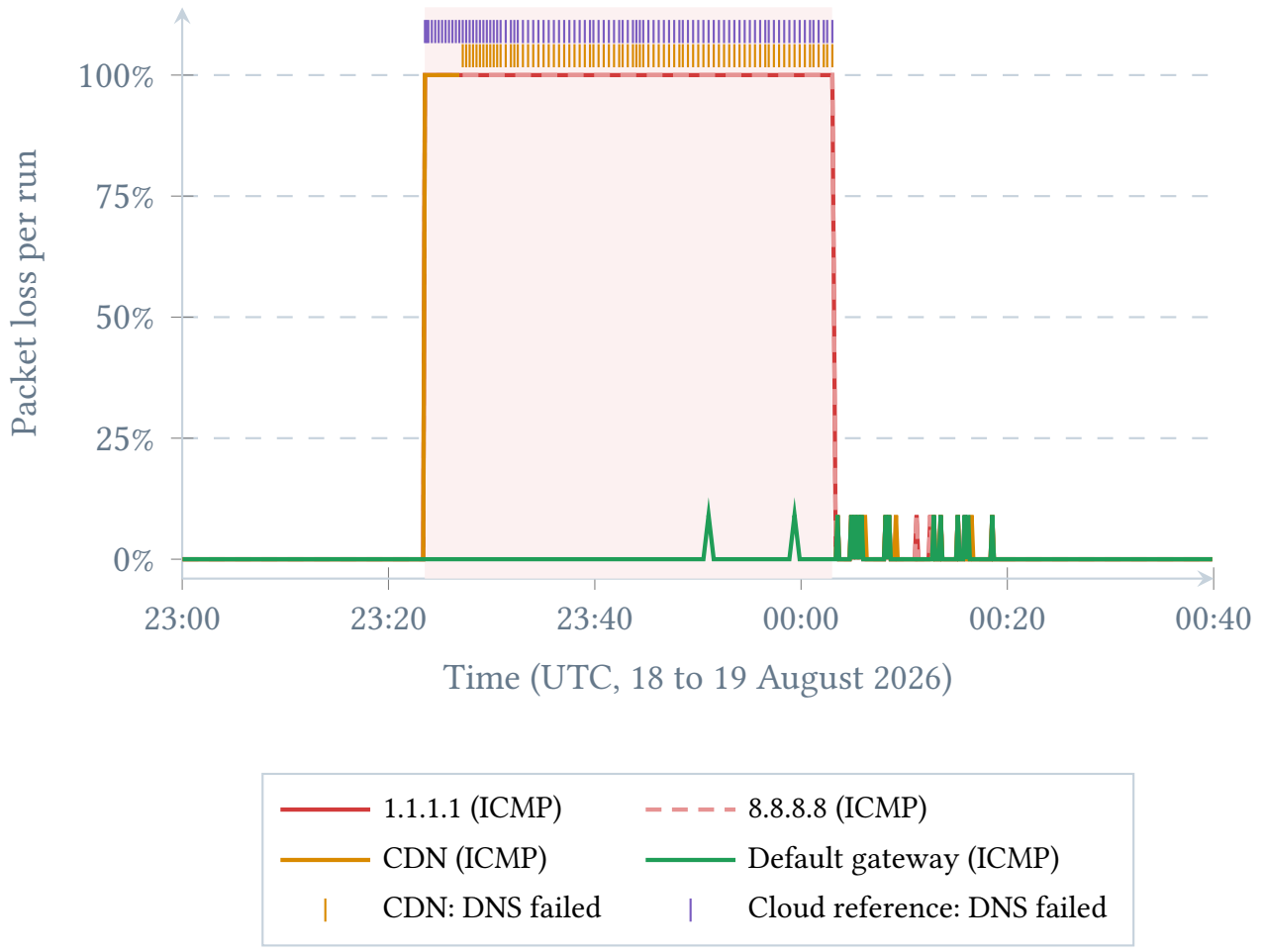
\begin{figure}[t]
\centering
\begin{tikzpicture}
\begin{axis}[
    width=\columnwidth, height=5.4cm,
    xmin=23.0, xmax=24.667,
    ymin=-4, ymax=114,
    xtick={23.0,23.3333,23.6667,24.0,24.3333,24.6667},
    xticklabels={23:00,23:20,23:40,00:00,00:20,00:40},
    ytick={0,25,50,75,100},
    yticklabel={\pgfmathprintnumber{\tick}\%},
    xlabel={Time (UTC, 18 to 19 August 2026)},
    ylabel={Packet loss per run},
    xlabel style={font=\footnotesize, text=figmute},
    ylabel style={font=\footnotesize, text=figmute},
    tick label style={font=\footnotesize, text=figmute},
    axis lines=left, axis line style={draw=figline},
    ymajorgrids, grid style={draw=figline, dashed},
    unbounded coords=jump,
    legend style={font=\scriptsize, draw=figline, at={(0.5,-0.34)},
                  anchor=north, legend columns=2,
                  /tikz/every even column/.append style={column sep=6pt}},
    legend cell align=left,
    clip=false,
]
\addplot[draw=none, fill=figbad!7, forget plot] coordinates
  {(23.392,-4) (24.050,-4) (24.050,114) (23.392,114)} \closedcycle;

\addplot[figbad, thick] table[col sep=comma, x=t_h, y=a1] {figures/outage-loss.csv};
\addlegendentry{1.1.1.1 (ICMP)}
\addplot[figbad!55!white, thick, densely dashed] table[col sep=comma, x=t_h, y=a8] {figures/outage-loss.csv};
\addlegendentry{8.8.8.8 (ICMP)}
\addplot[figreal, thick] table[col sep=comma, x=t_h, y=cdn] {figures/outage-loss.csv};
\addlegendentry{CDN (ICMP)}
\addplot[figgood, thick] table[col sep=comma, x=t_h, y=gw] {figures/outage-loss.csv};
\addlegendentry{Default gateway (ICMP)}
\addplot[only marks, mark=|, mark size=2.2pt, figreal] table[col sep=comma, x=t_h, y=cdn_fail_y] {figures/outage-loss.csv};
\addlegendentry{CDN: DNS failed}
\addplot[only marks, mark=|, mark size=2.2pt, figpath] table[col sep=comma, x=t_h, y=cloud_fail_y] {figures/outage-loss.csv};
\addlegendentry{Cloud reference: DNS failed}
\end{axis}
\end{tikzpicture}
\caption{Packet loss per baseline run by destination class, plotted from the
stored records, 23:00 to 00:40 UTC on 18 to 19 August 2026. Both
anchors returned no reply for 39 minutes (shaded) while the gateway kept
answering; the marks above the plot are cycles in which the CDN and
cloud hostnames failed at DNS resolution, so those classes have no loss
value there. A single end-to-end measurement would show only that the
network had failed.}
\Description{Line chart of packet loss per run against time from 23:00 to 00:40
UTC. The two anchors rise from zero to one hundred per cent at about
23:23 and return to zero at about 00:03, the CDN line is broken where
its DNS lookups failed, and the default gateway stays near zero
throughout with a few single-packet losses.}
\label{fig:outage}
\end{figure}

One fault in the month was neither planned nor staged. It was examined
afterwards, from the stored records. A scan of those records for one
pattern, minutes in which the default gateway answered normally while
every off-site destination failed completely, returned exactly one
episode over the period in which the gateway class existed, from
10:28 UTC on 18 August 2026: 18 August 2026 from 23:23 to 00:03 UTC,
39 minutes.\footnote{The dissertation gives 23:24 to 00:02 UTC from a
minute-level scan; the cycles with no reply run from 23:23:31 to 00:03:01
UTC.} Figure~\ref{fig:outage} shows it, plotted from the stored records.
In the 92 cycles in which both anchors returned no reply, the gateway
answered in every cycle and lost one packet in two of them; the CDN
hostname failed at DNS resolution in 80 cycles and got no reply in the
other 12; and the cloud class failed at resolution in all 92, so the
failure produced no TCP handshake measurement at all. The three kinds of
evidence are kept apart: ICMP runs with no reply, DNS failures, and the
gateway's replies. Together they show a loss of reachability beyond the
gateway as seen from this probe. That is consistent with a failure of
the building's uplink, which the probe cannot distinguish from a failure
nearer the router, and the event is called the uplink failure in this
paper with that qualification. The evidence shows that the gateway
stayed reachable, not that the wireless link was performing well: an
echo to the gateway says nothing about throughput or about other
devices. Because a run with no reply waits for \texttt{ping}'s deadline,
the scheduler skipped the cycles that overlapped, and consecutive cycles
were a median 30~s apart instead of 10~s.

\paragraph{What the system reported, then and afterwards.}
During the 39 minutes the dashboard had nothing new to show. The probe
could not reach the ingestion endpoint, and the 440 records timestamped
inside the window reached the database only after connectivity returned,
through the store-and-forward buffer, which was thereby exercised by a
real failure rather than a test. Nor would the code then deployed have
named the fault once the records arrived. Replayed through it, the
window read \emph{No measurements yet}, with the internet path showing
no data and a health score of zero: a run that failed and a run that
never happened were indistinguishable to the segment logic. The
\emph{down} and \emph{unstable} states of Section~\ref{sec:attribution}
were added on 20 August 2026 in response. Replaying the same stored
records through the corrected code gives \emph{Your WiFi is fine, but the
internet connection is down}, with the wireless link segment green and
the internet path segment red, and an ordinary verdict in the minutes
either side.

The result is therefore retrospective, and it is narrower than the
dashboard's sentence. The stored evidence separates a gateway that kept
answering from destinations that did not. The records of a tool
measuring only end to end could not do that, and it is the distinction a
resident or an operator needs. But the rule as deployed did not yet use
that evidence, and \emph{your WiFi is fine} claims more than a reachable
gateway shows. The scan also has limits. It could not see the earlier
episode of 11 August 2026, when the gateway class did not yet exist, and
the five-cycle episode of 28 August 2026, in which only ICMP was
dropped, does not match the pattern (both in
Section~\ref{sec:deployment}).

\paragraph{A second defect, found in the corrected version.}
Every reported figure was a median, and a median cannot represent a run
that produced no number. Viewed through the corrected code, the hour
from 23:45 to 00:45 UTC, which straddles the recovery, scored 93 out of
100, \emph{excellent}, while the same page reported that the internet
connection was down. Both statements were true, and together they made
no sense. The availability term of Section~\ref{sec:health} is the
result: the published score is quality multiplied by the share of
application tasks that completed. The same hour now reads 71.7,
\emph{fair}, from a quality of 94.4 and an availability of 76\%, 15 of
its 62 tasks having failed, and the hour ending 23:45 UTC reads 53
overall against 93 while working, because 27 of its 62 tasks did not
complete. A finished incident also fades from a trailing window: the
43 application runs that failed in the outage sit among more than eleven
thousand in a seven-day window around it, which read as fine. That is
why the dashboard's window can be made to end at any past instant.

\section{Fault injection}
\label{sec:faultinjection}

The outage of Section~\ref{sec:outage} is one kind of fault, a total
loss of off-site connectivity. This section tests the harder cases,
where a segment is degraded but still connected.

\paragraph{Method.}
Impairments were applied on the probe itself with \texttt{tc netem},
the Linux kernel's network emulator, which can delay, drop or slow
packets, here selected by destination address. Delaying every
destination except the default gateway gives the probe the symptoms of a
fault in the building's uplink; delaying only the real services gives
those of a fault at the third party. Three properties of the method
matter. First, the expected verdict for every scenario was written into
the scoring script, which was committed on 25 August 2026, three days
before the scenarios ran. Second, each scenario gave the probe its own
identifier, so its records are separate from the deployment's and from
one another and can be scored again at any time. Third, each scenario
ran for fifteen minutes with the application workloads speeded up to one
run per minute. The control showed the verdict settling after about
three minutes. It also showed that the first ninety seconds after a
restart are unreliable: the first page load includes starting the
browser (4,186~ms against a steady 3,706~ms), long enough to make the
control briefly report a fault that did not exist. Six of eight designed
scenarios were run. Scenario 07, a total loss of off-site connectivity,
and 08, an inbound queue deep enough to cause bufferbloat, were not,
because the site became unavailable at short notice; 07 is the fault
that occurred on 18 August.

One limit of the method should be stated plainly. Shaping by destination
on the probe changes what the probe's own packets experience; it does
not place a fault at a physical location. It reproduces the pattern that
a fault in a segment would produce across the destination classes, which
is what the attribution rule reads. It is therefore a test of whether
the rule responds to that pattern, not proof of where a fault is, and it
can distort measurements whose packets are addressed to a shaped
destination, as the first defect below shows. To claim that the tool
locates faults, impairments would be needed at physically separate
points: the client, a router under the experimenter's control and the
far end.

\begin{table}[t]
\caption{Fault injection results after the corrections described below.
Four of six scenarios are attributed as expected: the control and three
of the five impairments. \emph{Expected} and \emph{Reported} name the
segment the rule should and did implicate.}
\label{tab:faultmatrix}
\small
\begin{tabular}{@{}l>{\raggedright\arraybackslash}p{2.7cm}>{\raggedright\arraybackslash}p{1.5cm}>{\raggedright\arraybackslash}p{1.4cm}l@{}}
\toprule
\textbf{\#} & \textbf{Injected} & \textbf{Expected} & \textbf{Reported} & \textbf{Outcome} \\
\midrule
01 & nothing (control)                 & none          & none          & correct \\
02 & $+80$~ms, all destinations        & WiFi link     & WiFi link     & correct \\
03 & 5\% packet loss, all destinations & WiFi link     & none          & wrong \\
04 & inbound capped at 5~Mbit/s        & WiFi link     & WiFi link     & correct \\
05 & $+150$~ms, all except the gateway & internet path & third party   & wrong \\
06 & $+400$~ms, real services only     & third party   & third party   & correct \\
\bottomrule
\end{tabular}
\end{table}

\paragraph{Results.}
Table~\ref{tab:faultmatrix} gives the six scenarios. Scenario 02 raised the gateway round trip from
3.88 to 82.07~ms against 80~ms injected; 04 reduced throughput to
3.68~Mbit/s against a 5~Mbit/s cap; 05 raised the round trip to every
off-site destination to roughly 160~ms while the gateway stayed at
2.31~ms; 06 raised the real web target from 3,691 to 5,037~ms while the
cloud reference stayed at 216~ms. The impairments took effect; the
disagreements were in the analysis, which is what the exercise was for.

\paragraph{Three defects.}
Scored against the deployed implementation, only three of the six
scenarios were attributed correctly. The other three exposed three
distinct defects, two of which had gone unnoticed through a month of
live operation because the network was healthy enough that none changed
a verdict.

\emph{The link was measured by a contaminated proxy.} Scenario 05
impairs every destination except the gateway, so the local path is
unimpaired by construction and the tool should say so. It did not, because the link
verdict was derived from the first hop of the traceroute rather than
from the direct gateway probe. A traceroute packet is addressed to the
far destination and is only made to expire at the first router, which
then replies, so on its way out it has already been through any
impairment applied to that destination. Under scenario 05 the direct gateway probe read 2.31~ms
while the traceroute first hop read 153.25~ms for the same link at the
same moment. The tool had the correct measurement and used a distorted
one.

\emph{Packet loss was summarised by a statistic that cannot see it.}
Scenario 03 injects 5\% packet loss and changed no verdict. Of its 404 baseline runs, 279
reported exactly zero loss and the rest between 9\% and 23\%: with ten
packets per run, the probability that a run loses nothing is
$0.95^{10} \approx 0.60$, so the median across runs is zero, and the
health score, which used the median, saw a clean network. The mean over
the same runs is 3.76\%. The Grafana board already averaged loss over an
interval of time for this reason; the health score did not, so
two parts of one system disagreed about one measurement.

\emph{The reported cause could change with no change in the data.} Scenario 04 degrades
every endpoint at once and produced one vote for each of the three
segments. The implementation resolved ties with
\texttt{max(set(suspects), key=suspects.count)}, which returns whichever
element the set iterates first, and Python randomises string hashing per
process, so the same stored records could be attributed to a different
segment after a restart of the backend. Re-scoring identical data in
five separate processes produced three different answers. A diagnosis
that changes when nothing has changed cannot be checked or reproduced.

\paragraph{Corrections and re-scoring.}
The link verdict now prefers the direct gateway probe, falling back to
the traceroute first hop only where the gateway does not answer; packet
loss is aggregated by the mean of the per-run percentages, since the
packet-weighted rate of Section~\ref{sec:deployment} needs counters the
probe did not store; ties go to the segment nearest the user. Because
every record is stored, the corrections were checked on the same
records, without running the experiment again. This raises correct
attribution from three to four of six, with scenario 04 now passing by
rule, not by chance. Three automated tests reproduce the records that
exposed the defects. The corrected logic was evaluated
on the scenarios that exposed the defects, not on fresh ones, so the four
of six is a statement about those scenarios. The four are the control
and three of the five impairments, and each scenario ran once: the
figure describes these runs and is not an estimate of accuracy.

\paragraph{The two remaining failures.}
Neither is a coding error, and both limit what the tool can claim.
In scenario 03 no application headline metric crossed its configured
threshold: the page still loaded and the video still started. Against
the control no median came near a line: the real page load was 3,642~ms
against 3,692 (difference of medians $-49$~ms, bootstrap 95\% interval
$-98$ to $-4$, 15 runs against 30), cloud video startup 323 against
296~ms ($+27$, $-1$ to $+222$), the email fetch 561 against 554~ms
($+7$, $-41$ to $+92$), and the cloud download fell from 52.9 to
38.1~Mbit/s ($-14.8$, $-29.4$ to $+0.2$, on four runs against eight),
still well above its 20~Mbit/s line. The loss shows in single slow runs
instead. Two of the 15 real video startups took about 5.3~s, where the
slowest of the control's 30 took 0.35~s, and two of the 15 email fetches
took 3.2 and 10.5~s against a control maximum of 0.89~s: runs that
waited out a retransmission timeout. The gateway's mean loss was 4.99\%
against 0.54\% in the control.\footnote{Computed by
\texttt{notes/application\_results.py} from the exported application
records; each interval is a percentile bootstrap over 10,000 resamples
of the two groups. The control was started twice, at 21:00 and 21:23
UTC, and both runs are used, which gives a control median of 3,692~ms
where the dissertation's figure, quoted elsewhere in this section, is
3,691.} After
the correction the loss is visible
in the health score, whose responsiveness component falls from 88.5 to
58.8, but no segment is implicated, because attribution is driven by
workload thresholds and none was crossed. Loss influences the score
without influencing the diagnosis. Scenario 05 is more instructive. The
fault is on the internet path and the tool blamed the third party, as a
direct consequence of giving the real endpoint its own thresholds
(Section~\ref{sec:threeway}). The cloud reference is a small static page:
with 150~ms added it still loaded in 827~ms, inside its 2,000~ms line, so
the internet-path segment was never flagged. The real site is megabytes
across dozens of requests, so the same added latency pushed it from
3,691 to 4,693~ms, across its 4,000~ms line. The segment that crossed a
line first was not the segment at fault. A fixed threshold catches the
same proportional slowdown sooner on a heavy page than on a light one.
The remedy is to compare each endpoint with its own recent behaviour. The signal needed was already being recorded:
during scenario 05 the round trip to the cloud reference rose from about
35 to 183~ms, which the attribution logic never consulted.

\section{Eight measurement artefacts}
\label{sec:artefacts}

A measurement artefact is a result produced by the tool, not by the
network. Eight times the project reported a network problem that did not
exist, or missed one that did, because of a fault in the tool or its
analysis. They are collected here because together they make a point
that a list of separate bugs would not.

\begin{enumerate}
\item \textbf{A constant one-in-five packet loss from the ping
  invocation.}
  From the first day the baseline reported exactly 20\% loss to both
  anchors at every hour. The probe ran \texttt{ping -c 5 -w 5}; adding
  an explicit reply timeout (\texttt{-W 2}) and a deadline of twice the
  count removed it, and the rate fell to 0.33\%. The dissertation
  explained it as the 5~s deadline clipping the fifth reply, but with
  one-second spacing that reply is due at about 4~s, so the explanation
  does not fit, and the mechanism was never reproduced with the deployed
  iputils version. The observation and the fix stand; the cause is
  recorded as unresolved.
\item \textbf{Quantised packet loss drawn as continuous.} With five packets per
  run the reported loss is a multiple of 20\%. Of 8,634 samples, 8,490
  were exactly zero and 144 exactly 20\%, which drawn as points is a
  dense row of full-height bars. Averaging over a bucket recovers the
  rate, because equal packet counts per run make the mean of the
  percentages equal to packets lost over packets sent.
\item \textbf{An averaging interval too short for the average to work.}
  The dashboard averages the points in each interval of time before
  drawing them. At this density it chose an interval of about a minute,
  holding five or six runs, so one lost packet still swung an interval
  to five or ten per cent. A minimum interval of fifteen minutes, ninety
  runs, reduced the step to 0.22\%.
\item \textbf{A line drawn across absent data.} During the 19.88~h probe
  outage the time-series panels joined the points either side of the
  gap, rendering a day of missing data as a gentle drift in page load
  time. The panels now break the line on gaps longer than fifteen
  minutes.
\item \textbf{Medians blind to failed runs.} A window straddling the
  recovery from the uplink failure scored 93 out of 100 while reporting
  that the internet path was not responding (Section~\ref{sec:outage}).
\item \textbf{Medians blind to 5\% packet loss.} With ten packets per run, six
  runs in ten lose nothing, so the median across runs is zero and 5\%
  loss is invisible to it (Section~\ref{sec:faultinjection}).
\item \textbf{An ICMP parser reading at the wrong offset.} The iOS probe
  reported 100\% loss to every gateway on every network for three days,
  which was read as routers refusing echo requests. The routers were
  answering in about three milliseconds. On Darwin a datagram ICMP socket
  delivers the IPv4 header ahead of the ICMP header, and the parser read
  the reply type at offset zero, rejecting every real reply.
\item \textbf{A pooled decomposition that mixed periods.} The
  dissertation reported the gateway as 29\% of the median path to a
  public anchor and as the source of most of its variability. Its
  gateway median came from the samples after 18 August 2026 and its
  anchor median from the whole run, three quarters of it before that
  date, when the anchor was probed with five packets at one-second
  spacing and its median was 1.7~ms higher. In the same cycles the
  medians are 3.82 and 11.62~ms, and the p95/p50 contrast the variability
  claim rested on, 1.75 against 1.08, becomes 2.54 against 1.72
  (Section~\ref{sec:deployment}). 
\end{enumerate}

They share a shape. In each case a statistic, a chart, a parser or a
summary was applied to data it did not suit, and the result looked
exactly like a network fault. The ping invocation and the ICMP parser
were errors of collection, which stored false values; the quantised
bars, the averaging interval and the line across the gap were errors of
presentation over correct values; the two medians and the pooled
decomposition were errors of aggregation, the step that summarises many
values as one. None was found by a test. Each was found by questioning a number that
looked too regular: loss of exactly one packet in five at every hour,
total loss on every network, a score of 93 beside a red segment, a
decomposition whose two halves had different sample counts. Two of the
eight were found only by injecting faults, because a month of healthy
operation never let them change a verdict. The practical lesson is that
a measurement tool must be validated at both ends, collection and
presentation, because an error at either end produces the very fault the
tool claims to observe. Paxson's strategies for sound measurement, calibration
by examining outliers and spikes and by checks of
self-consistency~\cite{paxson2004sound}, describe how these were found.
Three further defects of the attribution logic, a failed run read as an
absent one (Section~\ref{sec:outage}), a contaminated link proxy and a
tie broken at random (Section~\ref{sec:faultinjection}), are errors of
diagnosis rather than of measurement and are not counted among the
eight.

\section{Related work}
\label{sec:related}

\paragraph{Measuring access networks from inside.}
Sundaresan et al.\ measured broadband links from gateways in thousands of
homes and showed that the last mile often dominates end-user latency and
that latency under load exposes oversized buffers on every ISP they
measured~\cite{sundaresan2011broadband}; BISmark is the testbed that
grew from it~\cite{sundaresan2014bismark}, and the SamKnows whiteboxes
behind regulator programmes such as Measuring Broadband America measure
the line continuously from a dedicated
device~\cite{fcc2011mba,samknows}. Those probes measure from the gateway rather than from a client on the
WiFi; BISmark's own paper includes wireless experiments and
application-level tests, so the difference is where the probe sits and
what is published, not what is measured. Later work from the same
group asked whether a slow download was limited by the home network or
the access link, from the gateway's vantage and for throughput
only~\cite{sundaresan2016lastmile}, and measured where web page loads
lose time in access networks~\cite{sundaresan2013web}. On the wireless
side, Patro et al.\ observed home WiFi experience from the access
point~\cite{patro2013observing}, Biswas et al.\ reported large-scale
measurements of WiFi behaviour from a cloud-managed access point
fleet~\cite{biswas2015meraki}, and two studies of operational
networks found the WiFi hop to be a large and variable share of
end-to-end round trip~\cite{pei2016wifi,sui2016wifi}. Netalyzr diagnosed
the edge network from a browser session~\cite{kreibich2010netalyzr}, and
HomeNet Profiler characterised home networks from a client
inside them~\cite{dicioccio2013homenet}. This paper's probe is a client
inside the WiFi, like the last two, runs application workloads rather
than network tests, and compares endpoints that share the access network
and differ in destination;
its paired baseline makes, for one network, the kind of comparison
those works make across many homes.

\paragraph{Platforms and consumer tests.}
Bajpai and Sch\"onw\"alder survey the measurement platforms, RIPE Atlas
and M-Lab among them~\cite{bajpai2015survey,ripe2015atlas,dovrolis2010mlab}.
Their probes have no local reference and run network-level tests rather
than a full browser, and they measure many sites shallowly where this
work measures one site deeply. Consumer speed tests~\cite{ookla}
typically report upload and download throughput from one run;
Cloudflare's also takes latency, jitter and packet loss and scores the
run for streaming, gaming and calls~\cite{cloudflare-aim}, and Feamster
and Livingood set out why a speed figure is hard to interpret and what
else should be measured~\cite{feamster2020speed}. Commercial endpoint
monitoring goes further. The ThousandEyes endpoint agent, for one, runs
scheduled network and web tests from the user's own machine and records
the wired or wireless connection, gateways, proxies and DNS servers
beside them~\cite{thousandeyes-endpoint}, which overlaps with what this
paper builds. No comparison with such products was made. What this paper
offers beside them is an open implementation and analysis, a reference
copy of the same content measured next to the real service, and an
attribution rule scored against a real fault and injected ones with its
failures reported.

\paragraph{Bufferbloat and responsiveness.}
Gettys and Nichols named the problem of over-full
buffers~\cite{gettys2012bufferbloat}; the IETF's responsiveness metric
measures latency under working conditions as a quality figure in its own
right~\cite{ietf-responsiveness}, and H{\o}iland-J{\o}rgensen et al.\
measured latency variation across the Internet from RIPE Atlas
probes~\cite{hoiland2016latency}. The probe's latency-under-load
workload is a much simpler measure in the same spirit and does not
implement the draft's procedure; the deployment's bufferbloat threshold
turned out to sit at the network's median (Section~\ref{sec:limitations}).

\paragraph{Path measurement and its artefacts.}
Augustin et al.\ showed how per-flow load balancing produces false links
and loops in traceroute~\cite{augustin2006paris}; Luckie et al.\ showed that
the choice of probe packet changes which hops and destinations a
traceroute observes~\cite{luckie2008traceroute}, and the probe uses TCP
probes to port 443 because some networks drop echo requests but pass
web traffic; Govindan and Paxson estimated the delay routers add in
generating ICMP time-exceeded replies and found few routers slow at
it~\cite{govindan2002icmp}, so the suffix minimum of
Section~\ref{sec:baseline} is a display heuristic for the occasional hop
that reports more than the hops beyond it, not a correction their
result calls for.
Sommers et al.\ showed how far probe-based loss estimates can sit from
the true rate~\cite{sommers2005loss}; the quantised-loss artefacts of
Section~\ref{sec:artefacts} are a cruder form of the same problem. RFC
2330 defines the metrics and the sampling discipline that the baseline
approximates~\cite{rfc2330}, and Paxson's strategies for sound
measurement~\cite{paxson2004sound} are the method the artefacts section
arrives at by experience.

\paragraph{Applications and thresholds.}
Wang et al.\ showed that computation is a large share of the browser's
critical path, so page load has to be measured in a
browser~\cite{wang2013wprof}; the probe reads the Navigation Timing
interface for that reason~\cite{w3c-navigation-timing}. Dobrian et al.\
found rebuffering to be the quality metric with the largest effect on
engagement~\cite{dobrian2011video}, and Krishnan and Sitaraman
established the causal direction with a quasi-experimental
design~\cite{krishnan2012video}; both underlie the video component's
weight. The thresholds and the score's anchoring draw on ITU-T G.1010,
G.1030 and P.800~\cite{itu-g1010,itu-g1030,itu-p800} as background for
engineering choices, and the share of traffic that is video on
Sandvine's annual report~\cite{sandvine2024}; none of these sources
validates the thresholds, the weights or the mapping from a measurement
to a score.

\section{Limitations}
\label{sec:limitations}

The limits of the study are the following.

\begin{itemize}
\item \textbf{One site, one month.} The deployment is a case study of
  one resident WiFi network over 32 days, and the paired analysis covers
  the 10.4 days after the five-class baseline began, 11 UTC days. The
  bootstrap intervals describe the uncertainty from which days were
  observed, not how other networks behave, and with 11 days they are
  rough. The probe's settings changed on 18 August 2026 (packets per
  run, spacing, parallel execution) at the moment the three classes were
  added and the probe restarted. The anchor medians moved across that
  instant (13.33 to 11.62~ms for 1.1.1.1, 12.04 to 10.70~ms for
  8.8.8.8), and the earlier data cannot say whether the network itself
  changed that day.
\item \textbf{Proxies, not direct measurements.} The gateway RTT
  includes the router's own handling of echo requests, which can be
  slower than its forwarding, and in 0.1\% of cycles a far-end RTT is
  below the gateway RTT. The TCP handshake RTT of the cloud reference
  includes connection setup at an Azure front end and is not directly
  comparable with ICMP echo.
\item \textbf{A vacation period.} The run fell in the university
  vacation, when student accommodation is at its emptiest. The
  dissertation's hour-of-day analysis found anchor latency varying by under one per cent between the quietest and busiest hours,
  which says more about when the measurement was taken than about
  student networks.
\item \textbf{Two-way, not three-way.} No local reference could be run
  at the site, so the comparison the design rests on ran with two of its
  three endpoints, and the wireless link was judged by the gateway probe,
  a proxy for the local path, with no like-for-like throughput or
  application-level measurement of the link alone.
\item \textbf{Fixed thresholds, set before the data.} The thresholds of
  Table~\ref{tab:thresholds} were engineering estimates. Against the
  deployment they are uneven: the real-site web line sits near the 95th
  percentile of observed loads, the download line was never crossed, and
  the 30~ms bufferbloat line sits within half a millisecond of the
  network's median added latency, so half of normal operation falls on
  the wrong side of it. They were deliberately not recalibrated after
  the fact.
\item \textbf{Absolute numbers not externally checked.} The planned
  comparison against the same tasks timed by hand beside the probe was
  not carried out, because the site became unavailable at short notice.
  The results show consistency and meaningful comparisons between
  endpoints, not that a reported page load is what a stopwatch would
  have shown.
\item \textbf{Short scenarios.} Each injected fault ran for fifteen
  minutes, so the comparison of the loss scenario with the control
  (Section~\ref{sec:faultinjection}) rests on 15 application runs
  against 30, and on three or four transfer runs: enough to show that
  no median crossed a threshold while single runs stalled, not to bound a
  small shift. Each scenario ran once, on an implementation then
  corrected against it, so the result describes these runs and is not an
  estimate of accuracy.
\item \textbf{Fault location was emulated.} The injected faults shaped
  the probe's own traffic by destination
  (Section~\ref{sec:faultinjection}). They reproduce a pattern of
  symptoms, not a fault at a physical place.
\item \textbf{Diagnosis by pattern is ambiguous.} Several causes can
  produce the same pattern of verdicts, the probe itself among them. A
  gateway that answers echo requests shows that the local path was up,
  not that throughput was normal or that other devices were served well,
  so the dashboard's \emph{your WiFi is fine} says more than the
  evidence does.
\item \textbf{Missing measurements depend on conditions.} When replies
  stop, runs wait for their time limit and the scheduler skips the
  cycles that overlap, so cycles in the uplink failure were 30~s apart
  instead of 10~s, and for 19.88~h the probe recorded nothing at all.
  Rates computed over produced cycles therefore under-weight bad
  periods, and availability is not the share of time the network was
  usable.
\item \textbf{The probe's own traffic.} The times at which the probe's
  own workloads ran were estimated from the schedule
  (Section~\ref{sec:deployment}). The stored application records would
  allow a direct alignment, which has not been done, and the probe's
  processor load, local queues and concurrent transfers could move
  several destinations together.
\item \textbf{Packet counts are reconstructed.} The probe stored the
  loss percentage that \texttt{ping} printed, and neither its counters
  nor its version, so the packets sent and lost are inferred from the
  percentages (Section~\ref{sec:deployment}). A later probe should store
  both.
\end{itemize}

\section{Conclusion and future work}
\label{sec:conclusion}

The project began from the gap between what a speed test reports and
what a person experiences, and built a probe that measures a WiFi
network as a client performing real tasks. Three findings stand out.
First, in paired measurements over 10.4 days the round trip to a public
anchor was a median of 7.75~ms above the gateway's, and the two were
slow together: when the gateway's was above its own 95th percentile, the
anchor's was above its own in 92\% of cycles, against 0.4\% otherwise.
This replaces the dissertation's pooled figures
(Section~\ref{sec:deployment}). Second, the records of a real 39-minute
loss of reachability beyond the gateway separate a gateway that kept
answering from destinations that did not. The rule as deployed would
have reported missing data; corrected, it reports a reachable local
network and a fault beyond it, consistent with a failure of the uplink.
Third, five injected impairments and a control, scored against verdicts
fixed in advance, exposed three more defects that a month of live
operation had not. After correction the control and three of the five
impairments are attributed as expected, on those same scenarios, and the
two failures are limits of fixed thresholds.

The lesson about method matters more than any of the three. Eight times
the tool reported a fault the network did not have, or missed one it
did, through a ping command, a chart, an averaging interval, a median, a
parser or a pooled summary, and no test caught any of the eight. A
measurement tool has to be validated at both ends, in how it collects
data and in how it presents them. One session of deliberate fault
injection found what a month of observation had missed.

\paragraph{Future work.}
Four improvements follow from the limits above. Attribution should
compare each endpoint with its own recent behaviour, not with a fixed
threshold, so that the same proportional slowdown is detected wherever
it occurs. The baseline should feed the segment verdict directly: during
scenario 05 it showed plainly that the internet path was slow, and the
rule never looked at it. The fault injection should be repeated on fresh
scenarios against a frozen implementation, and any claim about the
physical place of a fault needs impairments at separately controlled
client, router and server. And a site that permits a wired reference
would restore the third endpoint of the comparison, while several sites,
or one site in term and in vacation, would turn a case study into a
comparison.

A further direction is a newer type of network workload, the streamed
answers of AI chat services. This traffic differs from the workloads
measured here in three ways. An answer is one long response that arrives
in many small pieces, so there is no single moment at which it has
loaded. It needs very little bandwidth, so throughput says almost
nothing about it. And in ordinary conditions its pace is expected to be
set by the service that generates the text, not by the network. The
question is therefore not how fast the network is but, when an answer
pauses, what caused the pause: the access network, the path, the
service, a relay between them or the client all look the same to the
user. The approach of this paper is now being applied to that question:
endpoints are compared as patterns of evidence and never subtracted, and
a diagnosis is scored against faults placed at known points and may
answer that the evidence is ambiguous. That work is in progress and none
of its results are used here.

\paragraph{Code, data and ethics.}
The probe, backend and dashboards are public under the MIT licence at
the repository in Section~\ref{sec:intro}. The version of the code that
this paper describes is marked there with a fixed label, the git tag
\texttt{arxiv-v1}, so that it can still be identified after the code
changes. The analysis scripts written for this paper, and the
measurement records they read, are not in the repository. They are
available from the author on request, with details that identify the
site removed. The probe
measured only its own traffic, under the author's own resident account,
about 3.5~GB a day. It never captured other devices' traffic, and the
site appears in the records only as a label and a truncated hash of the
public address.

\begin{acks}
The author thanks Prof.\ Alan Marshall, who supervised the MSc project
this paper condenses and commented on the outline and a draft of this
paper.
\end{acks}

\label{refs-start}
\bibliographystyle{ACM-Reference-Format}
\bibliography{refs}

\end{document}